\documentclass[11pt,final]{amsart}
\theoremstyle{definition}
\newtheorem{definition}{Definition}[section]
\newtheorem{theorem}{Theorem}[section]

\newtheorem{algorithm}[theorem]{Algorithm}

\usepackage{url,upref,amscd}
\usepackage{balance}
\usepackage{cite}
\usepackage{fixltx2e}
\usepackage{graphicx}
\usepackage{amsaddr}
\usepackage{fullpage}
\begin{document}
\title[On $r$-adding walks]{On improvements of the $r$-adding walk in a finite field of characteristic 2}
\author{Ansari Abdullah}
\address{Interdisciplinary School Of Scientific Computing, Savitribai Phule Pune University, Pune, INDIA}
 \author{Hardik Gajera}
\address{Dhirubhai Ambani Institute of Information and Communication Technology, Gandhinagar,
Gujarat, INDIA}
 \author{Ayan Mahalanobis}
\address{Indian Institute of Science Education and Research Pune, Dr.~Homi Bhabha Road, Pune, INDIA}
\email{abdullah0096@gmail.com, kidrah123@gmail.com, ayan.mahalanobis@gmail.com}
\begin{abstract}It is currently known from the work of Shoup and Nechaev that a generic algorithm to solve the discrete logarithm problem in a group of prime order must have complexity at least $k\sqrt{N}$ where $N$ is the order of the group. In many collision search algorithms this complexity is achieved. So with generic algorithms one can only hope to make the $k$ smaller. This $k$ depends on the complexity of the iterative step in the generic algorithms. The $\sqrt{N}$ comes from the fact there is about $\sqrt{N}$ iterations before a collision. So if we can find ways that can reduce the amount of work in one iteration then that is of great interest and probably the only possible modification of a generic algorithm. The modified $r$-adding walk allegedly does just that. It claims to reduce the amount of work done in one iteration of the original $r$-adding  walk. In this paper we study this modified $r$-adding walk, we critically analyze it and we compare it with the original $r$-adding walk.
\end{abstract}
\thanks{The third author gratefully acknowledges the support of a NBHM research grant.}
\maketitle
\section{Introduction}
In a recent article in the Journal of Cryptology, Cheon et.~al.~\cite{korean} published a novel time-memory speed-up of the well known Teske's $r$-adding walk~\cite{teske1, teske2}. Cheon et.~al.~claim up to 10 times speed-up of the Pollard's rho algorithm to solve the discrete logarithm problem in a finite field. This has no doubt stunned the cryptography world. Ten times speed-up is a remarkable speed-up. Earlier, what would have taken ten years to solve a discrete logarithm will now take about one year. This type of claim must be verified and re-verified.

This paper is based on our work trying to verify that claim made by the authors~\cite{korean}. We \textbf{cannot support their claim} of ``ten times faster''. However, at best, we found the \textbf{modified $r$-adding walk}, the one proposed by Cheon et.~al., is about three to five times faster (depending on what $r$ one chooses) than the original $r$-adding walk proposed by Teske~\cite{teske1,teske2}. We would reiterate that we find the idea of the modified $r$-adding walk interesting and novel.

The only way we find to compare these two algorithms -- the $r$-adding walk~\cite{teske1,teske2} and the modified $r$-adding walk~\cite{korean} is to do an actually implementation of these two algorithms in an identical platform. In that implementation we must be very careful. The way we implement \emph{table look-up} and \emph{tag computation} will influence the outcome of our experiment. We choose Magma~\cite{magma} and C++ using the NTL library~\cite{NTL} as our language of choice. The reason for that is simple, Magma probably is the best language to manipulate polynomials and has one of the best available large finite field implementation. On the other hand, Cheon et. al.~ used NTL to do their experiments and we report on their findings.
In trying to implement the modified $r$-adding walk we implemented the algorithm many times with different design paradigms until we were reasonable certain that the implementation was optimal in speed. Then we implemented the original $r$-adding walk. We made sure that these two algorithms were as similar as possible. As a matter of fact we can speak with conviction that the only difference in these two implementations was, the \emph{field multiplication in the original $r$-adding walk was replaced by tag computation and table look-up}. We changed the $r$-adding walk to use the distinguished path segment to define its distinguished points and the same index function $\gamma$.

In the original and the modified $r$-adding walk we need an index function $\gamma$. We define the function $\gamma$ and have shown that this is one of the best choices possible (see Figure 2). This makes us confident about our findings. So the obvious question comes: why are our findings so different with that of Cheon et.~al.~\cite{korean}? Since the authors provide very little details about their implementation, we are unable to answer confidently. Moreover, we have serious issues with the use of $r=4,8$  in the modified $r$-adding walk. It is known and we have re-established in Section~\ref{sec1} that the choice for $r$ must be at least 16. This has substantial effect on the speed of the modified $r$-adding walk.  
\paragraph{\textbf{Acknowledgements}} We gratefully acknowledge the help of John Cannon with Magma~\cite{magma} and the magma team for providing us with a free license. We also thank Jayant Deshpande for stimulating conversations and Jung Hee Cheon for email correspondences. 
\section{The structure of the paper} This paper is organized in two parts. In the first part we study the $r$-adding walk. The rho length of a $r$-adding walk is defined to be the number of iterations in the walk before the first collision. In Section~\ref{sec1} we study the \textbf{distribution of the rho length}. Our study of the $r$-adding walk is different from previous studies as we are using the distinguished path segment to define the distinguished points. The purpose of this section is to (re)establish the fact that one should use large $r$ ($r\geq 16$). We thought that it is important to establish this fact because the modified $r$-adding walk uses $r=4,8$. They use the mean of the rho length for $r=4,8$ but doesn't compute the standard deviation. The variance is large, which makes their estimation ineffective.

We then explain the modified $r$-adding walk and many of its salient features in details in Section~\ref{sec2}. We then do the comparison with the original $r$-adding walk and produce our results (Tables~\ref{result1} \& 3). We developed a new table look-up method for Magma that makes the modified $r$-adding walk go faster.
\section{$r$-adding walk}\label{sec1}
Generic algorithms for solving the discrete logarithm problem are algorithms that do not use the structure or representation of the group. They use the operation of multiplication, inversion and equality in group elements. These kind of algorithms are restrictive by nature, they are often not the fastest algorithms. However, they are very powerful, they can be applied to the discrete logarithm problem in every possible scenario, be it the group of rational points of an elliptic curve or that of the group of units of a finite field. It is currently known that the complexity of solving the discrete logarithm problem in a finite cyclic group of prime order $N$, using any generic algorithm is at least $k\sqrt{N}$~\cite{shoup,nechaev}, where $k$ is a positive constant. So any new modification of a generic algorithm can make the $k$ smaller. This $k$ is tied to the amount of work done in one iteration.

The generic algorithm that we want to start our discussion with is the famous Pollard's rho algorithm. He first developed it to factor integers and then that was adapted to solve the discrete logarithm problem\cite{pollard}. The idea behind the Pollard's rho algorithm is simple, create an iterated random walk in a finite cyclic group. Since the set is finite, there will always be a collision in this random walk. From that collision find the logarithm. However, there is one problem with finding the collision, store all the elements of the random walk. Not only that, every time a new node of the walk is computed one must check that with all the previous nodes. This increases both the time and space complexity of the algorithm. Pollard found a clever solution to the problem. He introduced a function,
iteration by which will simulate a random walk. Let $G=\langle g\rangle$ be a group of prime order. We are given $g$ and $h=g^x$ where $x$ is the discrete logarithm. Pollard's function is as follows: 
\begin{center}
$
f(y)=\left\{
\begin{array}{lll}
gy &\text{if}&y\in G_1\\
y^2&\text{if}&y\in G_2\\
hy&\text{if}&y\in G_3
\end{array}\right.
$
\end{center}
where $G_1,G_2$ and $G_3$ is an almost equal sized partition of $G$. In this case the iterated random walk looks like  the Greek letter $\rho$ and so the name Pollard's rho algorithm. The rho structure indicates that once there is a collision, the walk will repeat itself. This changes the storage requirement dramatically as follows:  we pick a few arbitrary points, and call them \textbf{distinguished points}, we will only have to look for collision in those distinguished points. Another way to think
of distinguished points is laying traps. We lay a few  traps and hope that some of them will be on the repeating part of the $\rho$. When we have two elements in our trap, we know that there is a collision and the algorithm stops. It is clear that we won't catch the first collision this way, but we don't have to do the search either, and
the saving in space compensates for this increase in time. We should add here that Pollard didn't propose this distinguished point method. His idea was using the Floyd's cycle finding method. However we present Pollard's rho algorithm this way to motivate our next discussion.

Teske\cite{teske1,teske2} developed $r$-adding walk in the same spirit as the Pollard's rho algorithm (the spirit being the $\rho$, a repeating random walk) but in practice it works differently. Let $G$, $g$ and $h$ be the same as above. For some $r\in\mathbb{N}$, let $\{m_1,m_2,\ldots,m_r\}$ be a set of elements of $G$ of the form $g^\alpha h^\beta$ picked uniformly randomly where $\alpha,\beta$ are integers. This is usually done by choosing $\alpha$ and $\beta$ uniformly random. These $m_i$ will be referred to as \textbf{multipliers} in this paper.
Let $\gamma:G\rightarrow\{1,2,\ldots,r\}$ be a function. An $r$-adding walk $\mathcal{F}$ is defined iteratively as follows:
 \begin{equation}\label{eqn1}\mathcal{F}(Y)=Ym_{\gamma(Y)}.\end{equation}
Note that computation of a node in the $r$-adding walk requires one group multiplication and one evaluation of the function $\gamma$. 

The starting point of the $r$-adding walk is computed by choosing a positive integer $\alpha_0$ uniformly random from the set $\left\{1,2,\ldots,|G|\right\}$ and computing $Y_0=g^{\alpha_0}$. To compute $m_i$  two integers $\alpha_i$ and $\beta_i$ are chosen uniformly randomly from $\left\{1,2,\ldots,|G|\right\}$ and $m_i=g^{\alpha_i}h^{\beta_i}$. It is easy to notice that as the walk progresses, the nodes of the walk are of the form $g^\alpha h^\beta$ for positive integers $\alpha$ and $\beta$. When there is a collision we have $g^\alpha
h^\beta=g^{\alpha^\prime}h^{\beta^\prime}$. Which forms the equation 
\begin{equation}\label{eqn2}
\alpha+x\beta=\alpha^\prime+x\beta^\prime\mod |G|.
\end{equation}
Since $|G|$ is prime, this equation is easy to solve for the unknown $x$.  

Teske~\cite[Section 5]{teske2} has shown that for large enough $r$ and suitable $\gamma$ the $r$-adding walk simulates a \emph{random random walk}\footnote{A random random walk is a random walk with a random starting point. Random being chosen uniformly random.} very well. In this paper we are not repeating Teske's work. We are not looking at the randomness aspect of this $r$-adding walk. We want to study the distribution of the rho length of a $r$-adding walk.

Iterative walks depending on functions from a finite set to itself are extensively studied by statisticians, see~\cite{harris,eric,ross}. 
\begin{definition}[Iterated walk]
Let $\Omega$ be a finite set of size $n$ and $T:\Omega\rightarrow\Omega$ be a function. Let $x\in\Omega$. Then the iterative walk corresponding to $T$ is defined as $\left\{x_0,T(x_0),T^2(x_0),\ldots,\right\}$ where $T^k$ is defined as composition of
$T$ with itself $k$ times.
\end{definition} 
Let us define $S_T(x_0)=\left\{x_0,T(x_0),T^2(x_0),\ldots,T^k(x_0),\ldots\right\}$ then the
size of $S_T(x_0)$ is the number of steps required for the collision with $T$ starting from $x_0$. If we denote the size of $S_T(x_0)$ by $s$, then we are interested in the distribution of $s$ when $T$ is chosen uniformly random from the set of all functions.  

Let $X$ be a random variable that counts the number of steps before a collision, for an iterating walk from $T$, starting from an arbitrary element $x_0$. Harris~\cite[\S 3]{harris} has shown that the probability density function of $X$ converges to 
\begin{equation}
\Gamma(x)=x\exp^{-\frac{1}{2}x^2}, \;\; x>0
\end{equation}
 as $n$ tends to infinity, where
$x\sqrt{n}:=s$. This is the classic Rayleigh distribution and it is known that the mean $\mu(X) = 1$ and the standard deviation $s.d.(X)\approx 0.523$ in units of $\sqrt{\frac{\pi}{2}q}$. \footnote{The theoretical mean and standard deviation of Rayleigh distribution are generally in units of $\sqrt{q}$ but, our experimental mean and standard deviation are in units of $\sqrt{\frac{\pi}{2}q}$. So, we have provided the theoretical mean and standard deviation in units of $\sqrt{\frac{\pi}{2}q}$.} 
\paragraph{}
\textbf{Our question is}: 
Is the distribution of $X$ for the original $r$-adding walk close to the distribution function $\Gamma$? We answer this question in \textbf{affirmative for a large enough $r$} and a suitably chosen $\gamma$ by an experiment. Before we discuss our experiment, let us explain the methodology for the experiment. We have a probability density function $\Gamma$. We call the function $\Gamma(x)=x\exp^{-\frac{1}{2}x^2},\; x>0$ the test distribution function. We intend to generate data points (number of iterations before collision in a $r$-adding walk for different values of $r$) and then see if the data fits the test distribution function $\Gamma$. We will further plot the data and see the shape of that curve and compare that with the test distribution function. In judging, if a set of data fits a distribution -- the
\emph{quantile plots} (Q-Q plots) are very useful~\cite[\S4]{qqplot}. 
\begin{figure}
\centering
\includegraphics[trim=1cm 1cm 0cm 0cm, angle=0.1,width=12.6cm]{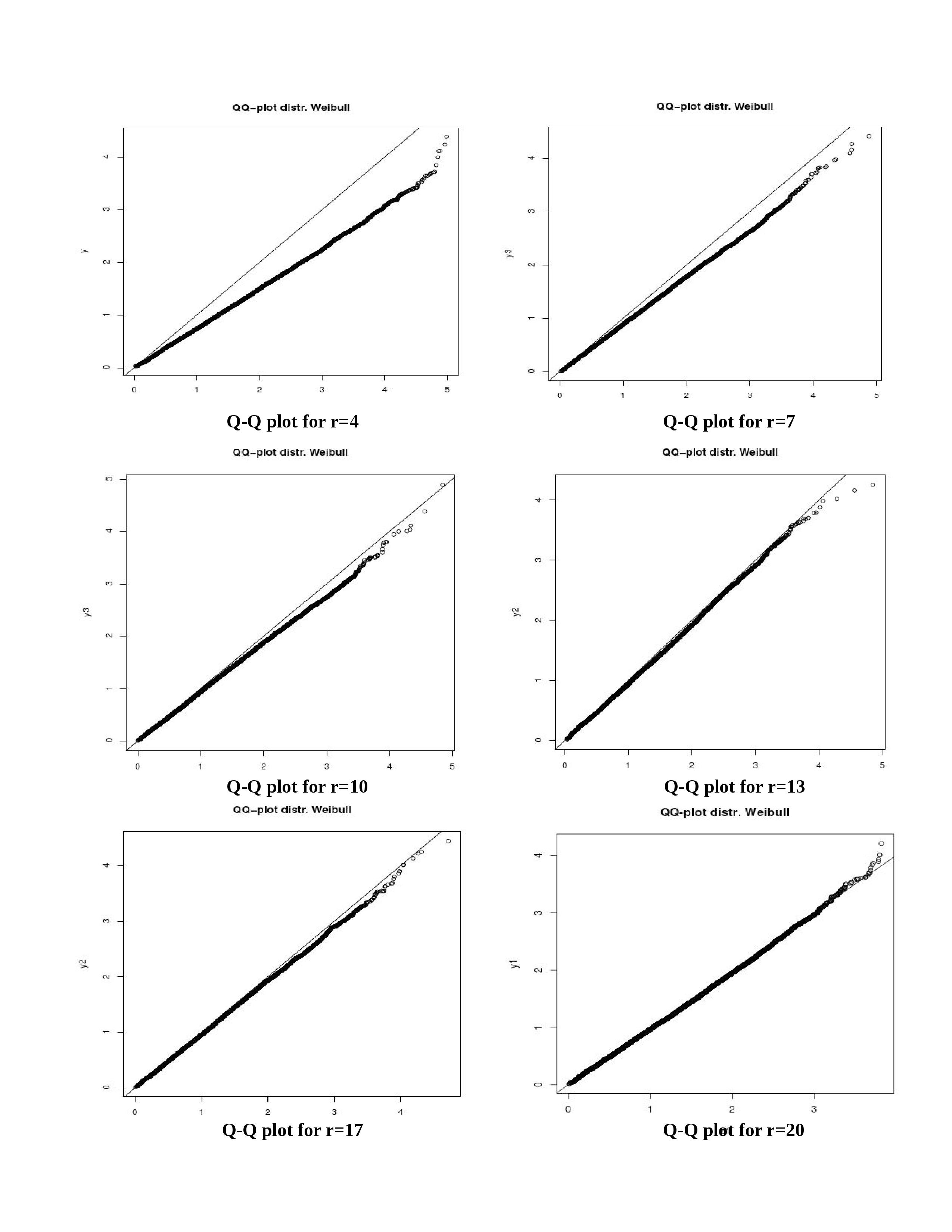}
\caption{The QQ-plots for average rho-length for different values of r over finite field GF$(2^{1023})$ and subgroup of size $2^{40}$.}\label{qq}
\end{figure}  
For our experiment, We have used a subgroup of the binary field $\mathbb{F}_{2^{1024}}$ of $40$-bit prime order. We have used the same $r$-adding walk described in~\cite{teske2}. We describe the index function\footnote{The index function is so chosen because we will use the same index function for the modified $r$-adding walk later.} $\gamma$. For each $r$, choose an integer $t\approx\log_2r$ and define the \textbf{tag function} $\tau:\mathbb{F}_{2^n}\longrightarrow\mathbb{F}_{2^t}$ where the image of $\tau$ is the vector of the coefficients of the first $t$ many highest degree terms in the polynomial representation of an element in $\mathbb{F}_{2^n}$. Define another function $\sigma: \mathbb{F}_{2^t}\longrightarrow \{1,2,\ldots,r\}$ where for each element $f(x)=a_0 + a_1x + \ldots + a_{t-1}x^{t-1}$  in $\mathbb{F}_{2^t},\sigma(f(x))= 1 + f(2) = 1 + a_02^0 + a_12^1 +\ldots+a_{t-1}2^{t-1}$. Note that for each $i$, $1\leq i \leq r, a_i$ is in $\mathbb{F}_2$. This means that $0\leq f(2) \leq r-1$ $(t\approx\log_2r)$ which implies that $2^t\approx r$. Then define $\gamma=\sigma \circ \tau$. Notice that $\tau$ is an additive function. It is clear from the above that once $\tau$ is computed, it is straightforward to compute the $\sigma$ and then $\gamma$. For the purpose of this paper one can use $\gamma$ and $\tau$ interchangeably. 

\begin{figure}
\centering
\scalebox{0.65}{\includegraphics[trim=4cm 5cm 3cm 3cm,width=15cm]{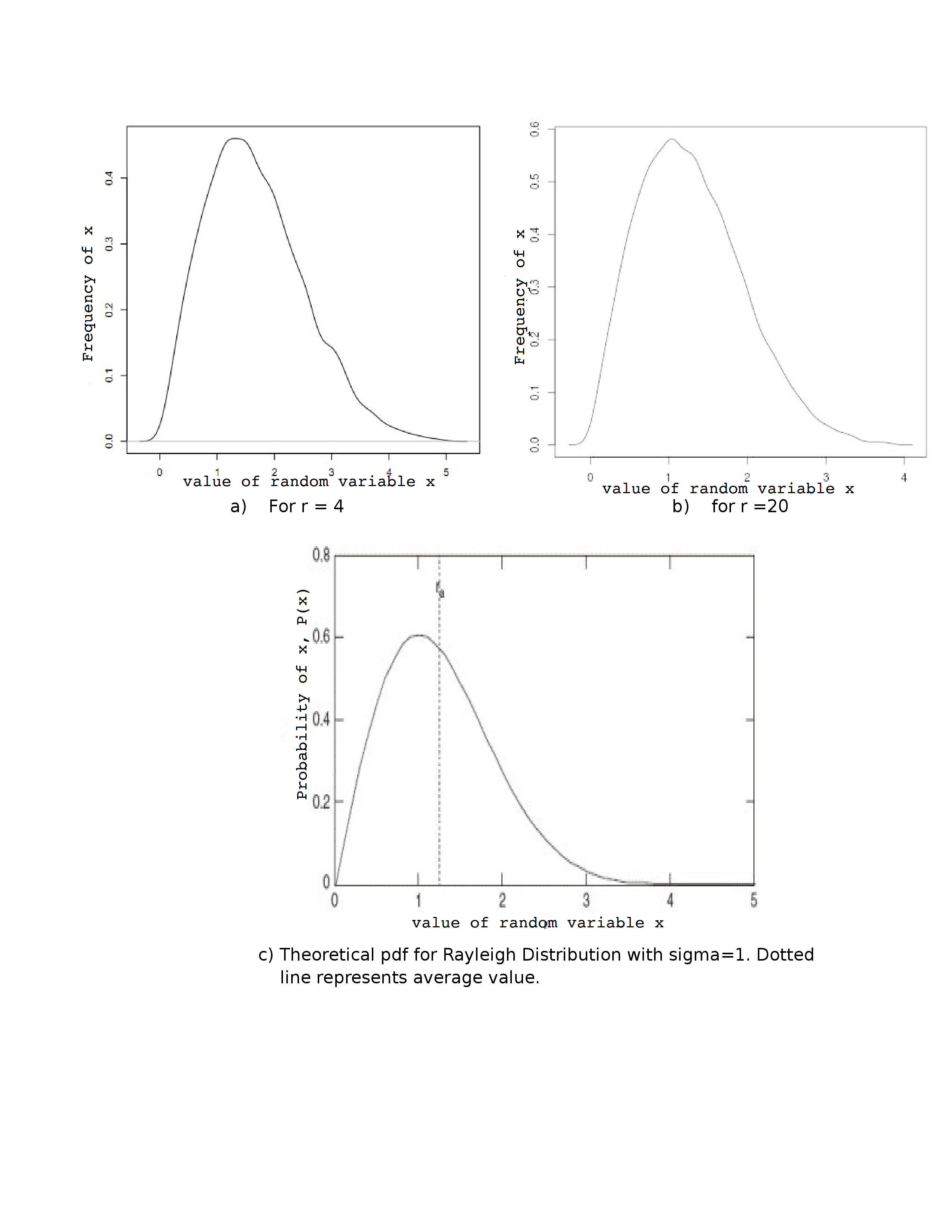}}	
\caption{Empirical Probability Density Function for two different r-values over finite field GF$(2^{1023})$ and prime order subgroup of size $2^{40}$.}\label{df}	
\end{figure}

We used the R software~\cite{r} to get the empirical probability density function and the Q-Q plots for our data. Figure~\ref{df} represents the empirical probability density function(epdf) of the variable $X$ for $r=4$ and $r=20$ over finite field $\mathbb{F}_{2^{1023}}$ and \textit{prime} subgroup of size $2^{40}$. Notice that the maximum $y$-value of the function for $r=4$ is less than $0.5$ and for $r=10$ it is greater than $0.55$. Theoretically, maximum $y$-value of  Rayleigh distribution is approximately $0.57$. Also, the epdf for $r=4$ is wider than the epdf for $r=20$. This means that for $r=20$, the probability of the value of $X$ lying in the neighborhood of the mean is higher than that for $r=4$. Hence, epdf for $r=20$ simulates Rayleigh distribution more closely than $r=4$. This means that $20$-adding walk is better than $4$-adding walk. Figure~\ref{qq} represents the Q-Q-plots for different $r$-values  over finite field $\mathbb{F}_{2^{1023}}$ and subgroup of size $2^{40}$. It seems clear from the Q-Q plots that as $r$ increases the distribution of $X$ comes closer and closer to the test distribution function $\Gamma$ and for $r=20$, it is virtually indistinguishable from $\Gamma$.
 Hence, one should use $r$ close to $20$ for practical purposes. Teske~\cite{teske2} also suggested that $r\geq16$ for practical purposes.
         
The \emph{rho length} of a $r$-adding walk is the number of iterations before the first collision in the walk. We ran tests to compare the average rho length of a $r$-adding walk for different values of $r$. Let us discuss in details our experiment. We have used cyclic subgroups of $G=\langle g \rangle$ of the field $\mathbb{F}_{2^{1024}}$. We have used four subgroups of prime order, they are $36$-bit, $37$-bit, $39$-bit, $40$-bit primes. For each DLP instances, an element $h\in G$ was chosen and a set of multipliers were randomly selected. Then the $r$-adding iteration function $\mathcal{F}$ was iterated from a random starting point until the first collision in the walk with the $\gamma$ defined previously. Once the first collision was found, the rho length was recorded. Note that we were looking for the first collision in a $r$-adding walk, so we didn't used the distinguished point method which gives approximate position of the collision. We have repeated this processes $10,000$ times for each of the four subgroups mentioned before. In the Table~\ref{rho}, the top row represents the size of the subgroup and all other rows represents average rho length for a specific $r$. The rho lengths are given in units of $\sqrt{\frac{\pi}{2}q}$. Let $\rho_r$ denotes the rho length of the $r$-adding walk in units of $\sqrt{\frac{\pi}{2}q}$. The table clearly shows that the $\rho_r$ is nearly stable and almost equal to one for subgroups of many different orders and for each $r\geq 16$. Hence, it is advisable to use $r \geq 16$ for all practical purpose. This reconfirms Teske's work~\cite{teske1, teske2}. We assume that the $\rho_r$ given in Table~\ref{rho} are roughly same even on very large prime order subgroup of $\mathbb{F}_{2^{1024}}$. Cheon et.~al.~\cite{korean} found similar results for $\rho_r$.  

\begin{table}
\caption{Average rho lengths for different $r$-adding walk over different subgroups of binary field $\mathbb{F}_{2^{1024}}$, in units of $\sqrt{\frac{\pi}{2}q}$ (the standard deviations of $\rho_r$ are given in the parentheses).} 
\centering
\footnotesize
\begin{tabular}{c c c c c c}
\hline
$r$-value  & $36$-bit & $37$-bit & $39$-bit & $40$-bit & average $\rho_r$\\
\hline
4 & 1.332(0.701) & 1.353(0.705) & 1.347(0.697) & 1.333(0.688) & 1.341 \\
5 & 1.212(0.627) & 1.187(0.620) & 1.189(0.627) & 1.191(0.620) & 1.195\\
6 & 1.146(0.598) & 1.137(0.583) & 1.131(0.588) & 1.133(0.585) & 1.137\\
7 & 1.127(0.575) & 1.113(0.575) & 1.108(0.570) & 1.121(0.585) & 1.116\\
8 & 1.087(0.566) & 1.092(0.565) & 1.084(0.570) & 1.081(0.567) & 1.086\\
9 & 1.069(0.557) & 1.077(0.565) & 1.077(0.561) & 1.072(0.560) & 1.074\\
10 & 1.080(0.563) & 1.062(0.556) & 1.064(0.558) & 1.075(0.558) & 1.070\\
11 & 1.055(0.547) & 1.081(0.549) & 1.063(0.547) & 1.053(0.544) & 1.063\\
12 & 1.051(0.545) & 1.049(0.554) & 1.046(0.542) & 1.053(0.550) & 1.050\\
13 & 1.060(0.543) & 1.080(0.551) & 1.038(0.550) & 1.050(0.533) & 1.057\\
14 & 1.059(0.546) & 1.057(0.548) & 1.053(0.546) & 1.059(0.547) & 1.057\\
15 & 1.061(0.547) & 1.084(0.566) & 1.032(0.540) & 1.033(0.537) & 1.052\\
16 & 1.037(0.541) & 1.037(0.542) & 1.044(0.541) & 1.034(0.540) & 1.038\\
17 & 1.034(0.548) & 1.038(0.559) & 1.038(0.540) & 1.040(0.545) & 1.037\\
18 & 1.038(0.542) & 1.035(0.538) & 1.024(0.535) & 1.040(0.544) & 1.034\\
19 & 1.028(0.543) & 1.035(0.538) & 1.027(0.537) & 1.027(0.537) & 1.029\\
20 & 1.033(0.538) & 1.031(0.538) & 1.017(0.536) & 1.020(0.521) & 1.025\\
\hline
\end{tabular}
\label{rho}
\end{table}
    
\section{Tag tracing -- a time-memory improvement of $r$-adding walk}\label{sec2}
Cheon et.~al.~\cite{korean} found an innovative way to speed up the
$r$-adding walk using a time-memory trade-off. Recall that one iteration
of the $r$-adding walk requires one field multiplication and one
evaluation of the function $\gamma$. The novel idea in this paper is not
to do the multiplication at every step, rather do it once in a
while. However multiplication is a binary operation, so it doesn't
matter how often one does that, it has the same number of
multiplications at the end. Authors circumvented this problem by storing various
products as group elements in a table $\mathcal{M}_l$ and calling them when
required. 

Recall that at any intermediate step, the iteration computes the product
$Ym_{i_1}m_{i_2}m_{i_3}\ldots m_{i_k}$ where $i_j\in\{1,2,\ldots,r\}$
and $k$ a positive integer.  If there is a table that can stores
the values of product $m_{i_1}m_{i_2}m_{i_3}\ldots m_{i_k}$ for various $k$, then these $k$
multiplications can be reduced to one multiplication. This is the
central idea that makes tag tracing go faster than the normal
$r$-adding walk. Select a positive integer $l$, and compute the
product after every $l$ steps using a table $\mathcal{M}_l$ that will be described
soon. 

There is one more thing that needs mention, the function
$\gamma:G\rightarrow\{1,2,\ldots,r\}$ can only be computed after the
product is computed. Recall that the iteration in $r$-adding walk is
$\mathcal{F}(Y)=Ym_{\gamma(Y)}$. So until $Y$ is available one cannot
compute $\gamma(Y)$ and the iteration cannot work. This gets in the way of the idea, ``multiply after every $l$
steps''. Authors~\cite{korean} solved this problem by introducing a
\textbf{tag} which is associated with every group element in the table ($\mathcal{M}_l$). Then $\gamma$ is a
function from this tag to $\{1,2,\ldots,r\}$. This involves a table
look-up in the modified $r$-adding walk that will slow things
down. We describe the algorithm of the modified $r$-adding walk in
details later. 

\subsection{Tag} Recall that we are working in a group
$G\subset\mathbb{F}_{2^\eta}^\times$. What we discuss will work for any
field extension of prime characteristic, however we do not know how to make this work outside of
finite fields.

In this case we represent the vector space $\mathbb{F}_{2^\eta}$ as a
$\eta$-dimensional vector space over the field of two elements
$\mathbb{F}_2$. We take the polynomial basis
$\{1,x,x^2,\ldots,x^{\eta-1}\}$ as the basis of $\mathbb{F}_{2^\eta}$
and any element $x\in\mathbb{F}_{2^\eta}$ can be written uniquely as a
polynomial of degree less than $\eta$ with coefficients over
$\mathbb{F}_2$. Fix a small positive integer $t$, the tag corresponding to $t$
for an element $f(x)=a_0+a_1x+\ldots,+a_{\eta-1}x^{\eta-1}$ in $\mathbb{F}_{2^\eta}$, 
is the coefficients of 
$\{x^{\eta-t},x^{\eta-t+1},\ldots,x^{\eta-1}\}$. In short, the tag is a binary
vector of length $t$ consisting of coefficients of the highest $t$ powers of the polynomial
$f(x)$. Notice that our polynomials are always of degree less than
$\eta$. One can also define the tag as an \textbf{additive function}
$\tau:\mathbb{F}_{2^\eta}\rightarrow \mathbb{F}_2^t$ where the image of
$\tau$ is the vector of coefficients of $t$ highest degree terms in
the polynomial basis
representation of an element in $\mathbb{F}_{2^\eta}$. It is easy to
verify that this tag is an additive function and respects scalar multiplication.
  
Now assume that $\tau\left(x^0m_i\right),\tau\left(xm_i\right),\tau\left(x^2m_i\right),\ldots,\tau\left(x^{\eta-1}m_i\right)$
is known, then we can compute
$\tau\left(fm_i\right)$ using the following formula:
\begin{equation*}
fm_i=a_0m_i+a_1\left(xm_i\right)+a_2\left(x^2m_i\right)+\ldots+a_{\eta-1}\left(x^{\eta-1}m_i\right)
\end{equation*}
and from the know properties of $\tau$ we have  
\begin{equation}\label{eqn4}
\tau(fm_i)=a_0\tau\left(m_i\right)+a_1\tau\left(xm_i\right)+a_2\tau\left(x^2m_i\right)+\ldots+a_{\eta-1}\tau\left(x^{\eta-1}m_i\right)
\end{equation}
The above statement follows from the distribution of multiplication
over addition in the polynomial algebra $\mathbb{F}_2[x]$ and is even true if we replace $m_i$
by a product $m_{i_1}m_{i_2}\ldots m_{i_k}$ for any positive integer $k$.
\subsection{The table $\mathcal{M}_l$} The table $\mathcal{M}_l$ has
$l$ rows. The first row contains $r$ cells. Each cell is numbered by
$\{1,2,\ldots,r\}$ corresponding to $\{m_1,m_2,\ldots,m_r\}$ stating from the left. 
The second row has $\binom{2+r-1}{2}=\binom{r+1}{2}$ cells. Total number of all possible $m_im_j$ where $1\leq i,j\leq r$ is $r^2$. Notice that we are in a abelian
group and $m_im_j=m_jm_i$. There are exactly $r$-many elements of the form $m_im_i$ where $1\leq i \leq r$. Hence, after removing duplicates, we are left with $\frac{r^2 -r}{2} + r = \frac{r^2+r}{2} = \binom{r+1}{2}$ many elements in second row. Continuing in this way we have the last row as the $l^{th}$ row. This row has $\binom{l+r-1}{l}$
cells. Each cell in the table corresponds to a vector
$(i_1,i_2,\ldots,i_k)$ for some positive integer $k$ and this vector
correspond to the group element $m_{i_1}m_{i_2}\ldots m_{i_k}$ where each $i_j\in\{1,2,\ldots,r\}$.

Each cell in the above table has four sets of information attached to
it:
\begin{description}
\item[Multiplier Information] This is a vector $(i_1,i_2,\ldots,i_k)$ of integers, where $k\leq l$ and $i_j\in\{1,2,\ldots,r\}$ for all $j$. We can assume that the vector is ordered. It contains the information on the multipliers involved in this cell.   
\item[Group element] The group element formed from multiplication of
  the multipliers involved in a cell, i.e., $m=m_{i_1}m_{i_2}\ldots
  m_{i_k}$ is computed and stored. 
\item [Exponent] Recall that each multiplier $m_i=g^{\alpha_i}
  h^{\beta_i}$ for some integers $\alpha_i$ and $\beta_i$. When these
  multipliers are multiplied, the exponents are added up in the
  product. This information $(\alpha,\beta)$ is the exponent, where
  $\alpha=\sum_{j}\alpha_{i_j}$ and $\beta=\sum_{j}\beta_{i_j}$. One
  needs the exponent information when the walk reaches a distinguished point.
\item[Tag] The vector
  $\left(\tau(m),\tau(xm),\ldots,\tau(x^{\eta-1}m)\right)$ is
  stored. 
\end{description} 
\subsection{An overview of the algorithm}
The modified $r$-adding walk proposed by Cheon et.~al.~\cite{korean} follows the
original $r$-adding walk closely. The only difference is, in the
modified one the multiplication is done after $l$ iterations and the
iteration uses a table look-up. In the
original $r$-adding walk multiplication is performed every iteration. 

Let $g$, $h$ and $m_i$ be as defined earlier. We compute the table
$\mathcal{M}_l$ as described above. Once that computation is done, we
start the iterated walk. An intermediate step in the iteration looks like
\[Y^\prime=Ym_{i_1}m_{i_2}\ldots m_{i_k}.\] Now we need to find
$\gamma(Y^\prime)=i_{k+1}$ where $\gamma$ is a index function from
$\mathbb{F}_2^t$ to $\{1,2,\ldots,r\}$. The function $\gamma$ was defined earlier. Assume that
$Y=y_0+y_1x+\ldots+y_{\eta-1}x^{\eta-1}$, and we know 
$(i_1,i_2,\ldots,i_k)$. The novel idea in the modified $r$-adding walk 
algorithm is, \textbf{we do not
have to compute the product $Y^\prime$, to find $i_{k+1}$}. Notice
that $(i_1,i_2,\ldots,i_k)$ is the multiplier information in the table
$\mathcal{M}_l$. Let us denote $m_{i_1}m_{i_2}\ldots m_{i_k}$ by
$m$. Let us warn the reader that this is just a notation to increase
readability not the product of $m_{i_1}m_{i_2}\ldots m_{i_k}$. Now we
do a \textbf{table look-up} and find the cell containing $(i_1,i_2,\ldots,i_k)$
as multiplier information. To that cell is attached the tag
\[\left(\tau(m), \tau\left(xm\right), \tau\left(x^2m\right),\ldots,\tau\left(x^{\eta-1}m\right)\right).\]

Now notice that,
\[Y^\prime=y_0m+y_1\left(xm\right)+y_2\left(x^2m\right)+\ldots+y_{\eta-1}\left(x^{\eta-1}m\right)\]
and from the additivity property of the tag function we have that
\begin{eqnarray}\label{eqn5}
i_{k+1}=\gamma\left(Y^\prime\right)=&y_0\tau\left(m\right)+y_1\tau\left(xm\right)+y_2\tau\left(x^2m\right)+\nonumber\\
&\ldots+y_{\eta-1}\tau\left(x^{\eta-1}m\right)
\end{eqnarray}
It is clear that the tag of $Y^\prime$ can be computed without computing
$Y^\prime$. So now $Y^{\prime\prime}$ can be determined  the same way
$Y^\prime$ was determined. We can continue this process $l$ times and
then compute the product from the pre-computed group element in the
table $\mathcal{M}_l$, that requires a table look-up. The full product is also computed when one
reaches a distinguished point. However that is a rare event and we
will totally ignore that. 
\subsubsection{Why is the modified $r$-adding walk faster?} In
computing an iteration in the original $r$-adding walk, we need to do
about $\eta^2$ multiplication in $\mathbb{F}_2$. We refer to
Equation~\ref{eqn5} to find the number of multiplications in
$\mathbb{F}_2$ for a single iteration in the modified $r$-adding
walk. Recall that $\tau\left(x^im\right)$ is already computed in
  $\mathcal{M}_l$ for each $i$ and is a vector of size $t$. So
  $\gamma\left(Y^\prime\right)$ can be computed in $t\eta$
  multiplications in $\mathbb{F}_2$. Since $t$ is significantly
  smaller than $\eta$, we have that $\eta^2\gg t\eta$ and an iteration in the modified walk
  is faster than the original $r$-adding walk. Recall that at every
  $l$ step there is a full product computation and assume that it
  takes $\eta^2$ multiplications. So to complete $l$ steps, in the
  modified walk we need $lt\eta+\eta^2$ filed multiplications compared
  to $l\eta^2$ multiplication in the original walk.  However, one has to do a table look-up and a tag computation as well and that can be time consuming.

\section{Comparison of modified $r$-adding walk with the original
  $r$-adding walk} We now describe the original $r$-adding walk by Teske and its modification by Cheon et.~al.~in more details. Cheon et.~al.~ described a novel idea of \emph{distinguished path segment} to find the distinguished point. We start with that.
  \begin{definition}[Distinguished Path Segment]
  Let $(g_i)_{i\geq 0}$ be a random walk over a finite group $G$. Let $\gamma : G \longrightarrow \{ 1,2,\ldots,r\}$ be an onto function. Fix a positive integer $\delta$. For $i\geq\delta-1$, the sequence $\{g_{i-\delta +1},g_{i -\delta +2},\ldots,g_i\}$ is called distinguished path segment if  
  \[\gamma(g_{i-\delta+1})=\gamma(g_{i-\delta+2})=\ldots=\gamma(g_i)=1 \hspace{10mm} \text{and} \hspace{10mm} \gamma(g_{i-\delta})\neq 1.\]
  \end{definition}
  Cheon et.~al.~\cite{korean} show that the expected number of function iterations before the appearance of the first distinguished path segment is  $\frac{r}{r-1}(r^\delta -1)$. So, the probability of a sequence $\{g_{i-\delta +1},g_{i -\delta +2},\ldots,g_i\}$ to be a distinguished path segment is $\frac{r}{(r-1)(r^\delta-1)}$. 
  
\subsection{$r$-adding walk}  
In this section, we briefly discuss the $r$-adding walk to solve the discrete logarithm problem. Throughout this section $G=\langle g \rangle$ denotes a finite cyclic group of prime order and $h=g^x$ where $x$ is the discrete logarithm. 

 Original $r$-adding walk requires an index function $\gamma: G \longrightarrow \{ 1,2,\ldots,r\}$. For our experiment, we had considered $\gamma = \sigma\circ\tau$ where $\tau$ is as discussed in Section 3. We want $\sigma$ to be a surjective function which is roughly pre-image uniform, i.e., the pre-image of each element is roughly the same size. Choose $t\approx log_2r$ and assign an unique non-negative integer less than $2^t$ as an image of an element of $\mathbb{F}_{2^t}$ under the map $\sigma$. This makes $\gamma$ pre-image uniform. For each $i=1,2,\ldots,r$ choose non-zero integers $1\leq\alpha_i,\beta_i < |G|$ and set multiplier $m_i$ equal to $g^{\alpha_i}h^{\beta_i}$. The $r$-adding iterating function $\mathcal{F}$ is as discussed in Section 2. Start iteration from $Y=g^{\alpha_0}$ where $\alpha_0$ is the random integer between $1$ and $|G|$. Compute $\mathcal{F}(Y)$ and set $Y=\mathcal{F}(Y)$. Since, $m_{\gamma(Y)}$ is of the form $g^{\alpha_{\gamma(Y)}}h^{\beta_{\gamma(Y)}}$, it is easy to keep track of the exponents of $g$ and $h$. Fix some positive integer $\delta > 0$ and define current element $Y$ as a distinguished point if it is the last element of a distinguished path segment of $r$-adding walk. The $r$-adding walk is travelled until the current element $Y$ is found to be a distinguished point. Whenever a distinguished point is reached, the current element is searched for in the table of distinguished points and is added to the table if it is not found. When there is a collision among distinguished points, we can use (\ref{eqn2}) to find the unknown $x$. 
  
  As discussed in Section 2, for a randomly chosen iteration function, the expected rho length is $1.253\sqrt{q}$ where $q$ is the size of the group. Since we are using distinguished points approach, one would expect to compute $1.253\sqrt{|G|} + \frac{r}{r-1}(r^\delta-1)$ iterations until a collision detection. Let MAX be the maximum number of distinguished points stored. Since the number of iterations until the appearance of the first distinguished point is $\frac{r}{r-1}(r^\delta-1)$, one should choose $\delta$ in such a way that MAX$\frac{r}{r-1}(r^\delta-1)\approx1.253\sqrt{|G|}$. At the same time one should keep in mind that $\frac{r}{r-1}(r^\delta-1)$ has to be much less than $1.253\sqrt{|G|}$.  
\subsection{Implementing modified $r$-adding walk in Magma}
On each iteration, the original $r$-adding walk computes a field multiplication whereas modified $r$-adding walk does not. Instead it requires a table look-up where the size of the table is quite large and a tag computation. As described in Algorithm 4.2, one needs to compute the table $\mathcal{M}_l$ before starting the algorithm. For a large enough group, time required to compute this table $\mathcal{M}_l$ is negligible compared to the time required to solve the discrete logarithm problem. 

\subsubsection{Tag computation}  We have described the concept of tag in details in Section 3.3. It is clear that faster the tag computation, faster the modified $r$-adding walk. In implementing our algorithms we are using Magma~\cite{magma}. One of the reasons we choose magma is that polynomial arithmetic and finite field implementation is the fastest in this package. We tried three different methods for this tag computation. We discuss those methods briefly.  Recall that $Y=y_0+y_1x+\ldots+y_{\eta-1}x^{\eta-1}$ and $m$ is $m_{i_1}m_{i_2}\ldots m_{i_k}$. We are using $m$ as a shorthand not the product. To compute $\tau\left(Ym\right)$, one needs to compute $y_0\tau(m)+y_1\tau\left(xm\right)+y_2\tau\left(x^2m\right)+\ldots+y_{\eta-1}\tau\left(x^{\eta-1}m\right)$.  The table $\mathcal{M}_l$ contains $\left(\tau(m), \tau(xm),\ldots,\tau(x^{\eta-1}m)\right)$. So we have to do this scalar multiplication and the addition.

\paragraph{\textbf{Method 1}}An obvious way to compute the tag is to loop over $y_i\tau(x^im)$ from $i=0\ldots\eta-1$. Advantage of this is that this will use polynomial arithmetic, which is fast, but the length of this loop will be equal $\eta$ which is $1023$ in our case. It turns out to be much slower than the later methods explained. 

\paragraph{\textbf{Method 2}}Another method is to use the in-built inner product function in Magma. Let $v$ be the vector of coefficients of the polynomial representation of $Y=y_0 + y_1x+ \ldots +y_{\eta-1} x^{\eta-1}$, i.e., $v=(y_0,y_1,\ldots ,y_{\eta-1})$ and from $\mathcal{M}_l$ we obtain $w=(\gamma(x^0m), \gamma(x^1m),\ldots,\gamma(x^{{\eta}-1}m))$. Then the inner product of $v$ and $w$ is $\gamma(Ym)$. This method won't require us to define a loop and at the same time, the inner product computation will use polynomial arithmetic. However, there is a serious disadvantage to this method. To use inner product, both the vectors $v$ and $w$ have to be in the same vector space. Note that the coefficients of $w$ are in $\mathbb{F}_{2^t}$, whereas the coefficients of $v$ are in $\mathbb{F}_2$. Hence, we need to \emph{coerce} (use an inbuilt embedding function in Magma) the vector $v$ into the vector space of the dimension $\eta$ over $\mathbb{F}_{2^t}$ and that makes it slower.

\paragraph{\textbf{Method 3}}The fastest tag computation that we could achieve was using $t$-many inner products instead of one. Again we were using the inbuilt Magma function for inner products. Recall that we need to compute $y_0\gamma(m)+y_1\gamma\left(xm\right)+y_2\gamma\left(x^2m\right)+\ldots+y_{\eta-1}\gamma\left(x^{\eta-1}m\right)$ and $\gamma(x^im)$ is a binary vector of size $t$. The idea is to compute these $t$ vectors in the sum independently. Each inner product has one input $\left(y_0,y_1,\ldots,y_{\eta-1}\right)$ and the other a vector of size $\eta$ of bits, where the $i\textsuperscript{th}$ entry comes from $\gamma(x^im)$. Which entry from $\gamma(x^im)$ gets chosen is decided by a loop. The first iteration of the loop uses the first entry of each $\gamma(x^im)$, the second the second entry from each $\gamma(x^im)$, and so on, the last entry in the $t\textsuperscript{th}$ iteration of the loop. So,
each loop gives the corresponding entry in the sum which is a binary vector of size $t$. For this we had to write an external loop in our program which runs for $t$ iterations. However, since $t$ is small this method was the fastest among all that we tried and was implemented.  

Here, the value of $l$ determines the speed up factor. If we increase the value of $l$, the number of consecutive iterations without product computation increases. One extreme value of $l$ is $\sqrt{|G|}$ where $|G|$ is the order of the group $G=\langle g\rangle$. However the size of the table $\mathcal{M}_l$ is $\binom{r+l}{r}$, increasing the value of $l$ increases the size of the table. Pre-computation time and the time required for a table look-up on each iteration increases with that. One needs to choose a value of $l$ in such a way that it balances the table look-up time and the storage availability.
    
\begin{algorithm}[Modified $r$-adding Walk]\mbox{}\\ 
\textbf{Input:}
\begin{description}
\item Field $\mathbb{F}_{2^{n}}$
\item Subgroup $G$=$\langle g\rangle$
\item An element $h$ of the subgroup $G$
\item Three positive elements $r$, $l$ and $t$.
\item $\mathcal{M}_l$
\end{description}
\textbf{The main algorithm}
\begin{enumerate}
\item Start with an empty table of distinguished points.
\item $(Y,\alpha,\beta,v) = \left(g^{\alpha_0},\alpha_0,0,(a_0,a_1,\ldots , a_{n-1})\right)$ where, $g^{\alpha_0} = a_0 + a_1x + \ldots + a_{n-1}x^{n-1}$.
\item \textbf{while} there are no duplicates among distinguished points
\begin{enumerate}
\item[] \textbf{do}$\left\{\begin{array}{l}
		 i=1\\
		 \text{\textbf{while} $i\leq l$ and $Y$ is not distinguished point}\\
		 \textbf{do}\left\{\begin{array}{l}
		 \text{set $m=mm_{s_i}m_{s_{i-1}} \ldots m_{s_0}$}\\
		 \text{look up for the vector $u=(\tau(m),\tau(xm),\tau(x^2m),\ldots ,$ }\\
		 \text{ $\tau(x^{n-1}m))$ in precomputed table $\mathcal{M}_l$}.\\
		 \text{compute $ss:=InnerProduct(v,u)$}\\
		 s_i=\sigma(ss).\\
		 i=i+1.\\
		 \end{array}\right.\\
		 (Y,\alpha,\beta,v)=(Ym,\alpha+\alpha_m,\beta+\beta_m,(c_0,c_1,\ldots ,c_{n-1}))\\
		 \text{where, $Ym=c_0 + c_1x + c_2x^2 + \ldots + c_{n-1}x^{n-1}$}.\\	
		 \text{\textbf{if} Y is a distinguished point}\\
		 \text{\textbf{then} add $(Y,\alpha,\beta)$ to the table of distinguished points.}\\
		 \end{array}\right.$ 
\end{enumerate}
\item[] Solve DLP using exponents $\alpha$ and $\beta$ of duplicate elements and Equation ($2$).
\end{enumerate}
\end{algorithm} 
  
\subsection{Table look-up}
In the modified $r$-adding walk, the multiplication in field $\mathbb{F}_{2^\eta}$ is replaced by few multiplications in $\mathbb{F}_2$ (tag computation) and a table look-up.  The size of the table $\mathcal{M}_l$ is $\binom{l+r}{r}$. For $l=10$ and $r=4,8,16,20$, the size of the table is $\binom{10+4}{4}\approx2^{10}$, $\binom{10+8}{8}\approx2^{15.4}$, $\binom{10+16}{16}\approx2^{22.3}$ and $\binom{10+20}{20}\approx2^{24.8}$, respectively. We can not ignore the time required for table look-up in the modified $r$-adding walk because the size of the table large. So during implementing the modified $r$-adding walk algorithm, we need to use the most efficient way for  table look-up. 

In general, the best method for table look-up is the binary search, but we came up with something even better. So, let us count the number of basic operations required for the table look-up in the binary search method.
\subsubsection{Binary search method} Let, the size of the table be approximately equal to $2^w$. Then on each iteration, binary search method will require $w$ steps of vector comparisons. Since the length of the vector $(i_1,i_2,...,i_k)$ varies from $1$ to $l$, each vector comparison requires on an average $l/2$ integer comparisons. Hence, the binary search method requires total $lw/2$ many integer comparisons on each iteration. For large value of $l$ and $r$, $lw/2$ is significantly large. For example, for $l=10,r=16$, $lw/2\approx110$. It takes a significant number of steps even though we are using the best known method. We tried many other approaches as well including the inbuilt index search algorithm in Magma. Finally, we developed our own algorithm using some pre-computation. 

\subsubsection{Pre-computation method} 
Note that $m$ in the table $\mathcal{M}_l$ is of the form $m_1^{a_1}m_2^{a_2}\ldots m_r^{a_r}$. We can pre-compute the vector $V$ of length $(l+1)^r$ whose element $V[a_1(l+1)^{r-1} + a_2(l+1)^{r-2}+\ldots + a_r]$ store the position $i$ of the element $m=m_1^{a_1}m_2^{a_2}\ldots m_r^{a_r}$. Note that to find the position of  $m=m_1^{a_1}m_2^{a_2}\ldots m_r^{a_r}$, we need to look at $V[a_1(l+1)^{r-1} + a_2(l+1)^{r-2}+\ldots + a_r]$. Let, $s = a_1(l+1)^{r-1} + a_2(l+1)^{r-2}+\ldots + a_r$. For any given $1\leq i \leq r$, to find the position of $m'=mm_i$, we need to look at $V[s+(l+1)^{r-i}]$. We can also store this $s$ of previous multiplier $m$ and  $(l+1)^i$ for each $0\leq i \leq (r-1)$. Hence, on each iteration, the table look up requires just one integer addition. We know that integer addition requires around $N$ steps where $N$ is the maximum of the number of digits of integers to be multiplied or added. For $l=10,r=16$, $\binom{l+r-1}{r}=\binom{25}{16}\approx2^{20.9}\approx 10^6$. Hence, for $r=16,l=10$, this method requires around $6$ basic steps on each iteration but, look at the size of the vector $V$. If we assume that each element of $V$ consumes $1$-bit of space, then for $r=16,l=10$, the vector $v$ consumes $11^{16}$ bits of space and $11^{16}$ bits $> 5 \times 10^6$ GB which is a huge space. Hence, this method is feasible only for small value of $r$ and we have used this method for $r=4,l=10$. We need to use different method for $r=16,l=10$.  Let us describe the another method for large value of $r$.

Let, $Y$ be the last fully computed element and $m=m_{i_1}m_{i_2}\ldots m_{i_k}$ be the last multiple. The index of $m$ in the table $\mathcal{M}_l$ is $\{i_1,i_2,\ldots,i_k\}$ and we denote it by $\text{ind}\textsubscript{m}$. Let $\gamma(Ym)$ be $i_{k+1}$. This means the next element on the walk is $Ymm_{i_{k+1}}$. Set $m^\prime=mm_{i_{k+1}}$. We need to find the index of the $m^\prime$. Note that given an element $m=m_{i_1}m_{i_2}\ldots m_{i_k}$ and a multiplier $m_{i_{k+1}}$, there is an unique element $m^\prime=mm_{i_{k+1}}$ in the table $\mathcal{M}_l$. We can store the index of $m^\prime$ for each possible pair $(m,m_{k+1})$. Note that if the multiple $m$ is of the form $m_{i_1}m_{i_2}\ldots m_{i_l}$, i.e., $m$ is the combination of the $l$-many multipliers, then we compute the $\gamma$-value of the next element by computing full product.  So, we are not considering pairs of the form $(m,m_{k+1})$ where $m=m_{i_1}m_{i_2}\ldots m_{i_l}$. We are left with $\binom{l+r-1}{r}$ many multiples. For each multiple $m$, we have $r$ many choice for $m_{k+1}$. Hence, total number of pairs of the form $(m,m_{k+1})$ is $r\binom{l+r-1}{r}$. We pre-compute a vector $V$ of length $r\binom{l+r-1}{r}$. We divide $V$ in $r$ equal parts of length $\binom{l+r-1}{r}$ where each part corresponds to a particular multiplier $m_{k+1}$. For example, first $\binom{l+r-1}{r}$ many entries corresponds to the pairs of the form $(m,m_1)$. If index of $m$ is $\text{ind}\textsubscript{m}$, then we store the index of $m'=mm_{k+1}$ in the table $\mathcal{M}_l$ as the $\left(k\binom{l+r-1}{r}+\text{ind}\textsubscript{m}\right)\textsuperscript{th}$ entry of $V$. On each iteration, we can find the index of $m'=mm_{k+1}$ using the index of $m$ and $i_{k+1}$. So the index of $m^\prime$ is $V[(s-1)\binom{l+r-1}{r}+\text{ind}\textsubscript{m}]$ where $\text{ind}\textsubscript{m}$ is the index of $m$ in the table $\mathcal{M}_l$. Note that we need to compute $\binom{l+r-1}{r}$ once because $l$ and $r$ are fixed. Hence, on each iteration this method requires one integer multiplication and one integer addition. We know that integer multiplication requires around $N^2$ steps and integer addition requires around $N$ steps where $N$ is the maximum of the number of digits of integers to be multiplied or added. For $l=10,r=16$, $\binom{l+r-1}{r}=\binom{25}{16}\approx2^{20.9}\approx 10^6$. Hence, for $r=16,l=10$, this method requires around $6^2 + 6 = 42$ basic steps on each iteration. In Section 5.3.1, we have shown that the binary search method requires $110$ basic steps for the same parameter. So, this method is faster than the binary search and we have used this method in our experiment for $r=16,l=10$.

\subsection{Results of our experiments in Magma}\mbox{}\\

Theoretically, it seems that the modified $r$-adding walk is faster than the original $r$-adding walk. To answer, how fast is the modified $r$-adding walk compared to original $r$-adding walk, we have tested both these algorithms on the prime order subgroup of the binary field $\mathbb{F}_{2^\eta}$. 

\subsubsection{Speed Comparison}  
We want to compare the time required for the original $r$-adding walk and the modified $r$-adding walk to solve the discrete logarithm problem in the same group. If we aim for practical parameters, i.e., DLP in a subgroup of size a $80$-bit prime, then it won't be possible to solve the DLP. So, we have measured the speed of the first few iteration of the walk for large parameters. One can find the time required to solve a  DLP for each walk in a large group using the average speed of each iteration and the expected rho length.  Let $v$ be the speed of each iteration and $L$ be the expected rho length, then the a DLP can be solved in time $v\times L$. As the results should not be biased toward modified $r$-adding walk, we have implemented both the method on same platform using Magma\cite{magma}. We use the binary field arithmetic functions from the Magma.

Let us explain the details of the experiment. We have selected a cyclic group $\langle g\rangle\subseteq\mathbb{F}_{2^{1023}}$ of order a $47$-bit prime. All other parameters were chosen as described in Section 4. Both the algorithms are same as far as possible. We ran the $20$-adding walk algorithm and the modified $r$-adding walk for different values of $r$ and $l$. Timing was started after the full computation of the respective multiplication tables $\mathcal{M}$ and $\mathcal{M}_l$. For each parameter set, we ran $100$ different DLP instances with $100$ tests for each instance. We measured the time of the first $10^7$ iterations in each case. Table~\ref{result1} provides the time require for $10^7$ iterations by both the algorithms. Here, we have excluded the pre-computation time for modified $r$-adding walk because time required for pre-computation is negligible compare to the time required to solve a discrete logarithm problem in the large group. 

\begin{table} 
\caption{Average time required for $10^7$ iterations of various methods on $47$-bit prime order subgroup of $\mathbb{F}_{2^{1023}}$ and the average rho length as in Table~\ref{rho} using Magma.}
\centering
\begin{tabular}{c c c c c c}
\hline
r & l & t & time & $\rho_r$\\
\hline
20, original & - & 6 & 121.37 & 1.025 \\
4, modified & 10 & 2 & 20.68 & 1.341 \\
16, modified & 10 & 4 & 51.6 & 1.038 \\
\hline
\end{tabular}
\label{result1}
\end{table}

From Table~\ref{result1}, the average ratio to solve a  DLP  by original and modified $r$-adding walk with $r=4$ is $\frac{124.404}{27.718}\approx 4.49$ and for $r=16$ is $\frac{124.404}{53.415}\approx2.33$.    

\subsection{Results of our experiments in C++ using the NTL library}
In the original $r$-adding walk there is one field multiplication in each iteration apart from evaluating the function $\gamma$. 
In the modified $r$-adding walk, one field multiplication is
done after $l$ steps and each one of the $l$ steps consists of computing the image of $\gamma$ which is the tag and a
table look up. Thus if the 
modified $r$-adding walk is executed for $n$ iterations actual field multiplication is done $n/l$ times. Number of field
multiplication done in the modified walk depends on $l$. When $l$ = 10,  the ratio of the work done for the actual multiplication 
in modified walk to actual multiplication step in original walk 
is ten times less. In other words, the original walk uses $10$ times more field multiplication than the modified walk. 
In both the algorithms the number
of evaluations of the function $\gamma$ remains the same. Table look up and tag computation are two extra steps in the modified walk. 
Table look up can be optimized as discussed earlier to get required information in constant 
time.

In our C++ implementation, the table look up was done using the binary search method.
Tag computation becomes a bottle neck as the time spent in this operation in a single iteration is the most dominant step.
The inner product function provided by NTL cannot be used for tag computation directly.
One method that could be used for tag computation is matrix multiplication,
i.e, representing field 
element and tag vector as matrices and multiplying these matrices. We use the same notations from Section 5.2.1.
Let $Y=y_0+y_1x+\ldots+y_{\eta-1}x^{\eta-1}$.

The \emph{matrix} $v$ is a 
$1 \times \eta$ matrix,
\[ v = \left (\begin{array}{cccc}
y_0, & y_1, &  \dots ,&y_{\eta-1} \end{array}\right).\] 

Tag vectors $\gamma(x^im)$, $i=0,1,\ldots,\eta-1$ are used as rows to construct the $\eta \times t$ matrix $w$, 

\[ w = \left (\begin{array}{cccc}
a_{01} & a_{02}  &  ... &a_{0t} \\
a_{11} & a_{12}  &  ... &a_{1t} \\
\vdots & \vdots & \ddots &\vdots \\
a_{\eta-1 1} & a_{\eta-1 2} &  ... &a_{\eta-1 t} \end{array}\right).\]

Then $v\times w$ is the required tag, which is the input to the function $\gamma$.
However, we found this method to be not practically useful as matrix multiplication was slow. Furthermore, NTL does not allow access to elements of GF2E as elements of an array. Thus
we have to "type cast" often to make $v$ a vector. 

Note that elements of $v$ are either $0$ or $1$. Using that, 
our implementation of tag computation was done by iterating over all non-zero elements 
in each column of $w$ and compute the sum corresponding to non zero elements in $v$. In practice, we didn't create the matrix $w$ explicitly but worked with the tag vector in the table $\mathcal{M}_l$ directly.
This was our tag computation for the experiments.

\subsubsection{Sparse vs.~ normal irreducible} In this section we show that the comparison of the modified walk and the original walk depends on the irreducible polynomial used to define the finite field $\mathbb{F}_{2^{\eta}}$. To summarize our result in C++, we found, for an arbitrary irreducible polynomial the modified walk for $r=4$ is about 1.5 times faster than the original walk.
If a sparse\footnote{In Magma we didn't use sparse irreducible polynomials.}  irreducible polynomial is used, computation time taken by both algorithms is about the same. For $r=16$, the original with sparse irreducible polynomial is actually faster and with arbitrary polynomial both the algorithms take about the same time, see Table 3 below.

In the modified $r$-adding walk the intermediate steps -- evaluating the $\gamma$ function,
computing the tag and the table look up, are almost independent of the field operation. On the other hand, in the original $r$-adding walk, there is one field multiplication in each iteration. Thus an arbitrary irreducible polynomial makes the original walk slower.
\begin{table}[ht]
\begin{tabular}{ l|c|c|c|c| }
  \hline
  \multicolumn{1}{c|}{$r$, $l$} 
& \multicolumn{2}{|c|}{Original}
& \multicolumn{2}{|c|}{Modified}
\\
  \hline
  & Sparse & Arbitrary & Sparse &Arbitrary \\
\hline
$4$, $10$ & $35.35$ & $82.23$ & $50.45(38.67)$ & $53.13(40.70)$\\
\hline
$16$, $10$ & $40.24$ & $87.89$ & $62.36(50.12)$ & $71.09(46.26)$ \\ 
\hline
\end{tabular}
\caption{Average time in seconds for 100 instances where each instance performs $10^7$ iterations for original and modified $r$-adding 
walk on 47-bit prime order subgroup of $\mathbb{F}_{2^{1023}}$ using the NTL library in C++. Time for tag computation given in parentheses.}
\end{table}
\section{Conclusion}
Cheon et.~al.~\cite{korean} used binary field arithmetic functions from NTL library and measured the average time required by both algorithms for $10^8$ iterations on $206$-bit prime order subgroup of the binary field $\mathbb{F}_{2^{1024}}$. For $r=4, l=10$, they found that modified $r$-adding walk is around $8.6$ times faster than the original $r$-adding walk. Our results are different from their results. In our case, using magma, the original $20$-adding walk took on an average $121.37$ seconds for $10^7$ iterations while in ~\cite{korean}, the average time for original $20$-adding walk is around $406$ seconds per $10^8$ iterations. In our implementation, there are three main steps. Field multiplication, conversion of field element into a vector and tag computation.  If we ignore the $\gamma$ function, our time is around $40.4$ seconds for $10^7$ iterations which is around $404$ seconds for $10^8$ iterations. This is nearly equal to the respective claim in ~\cite{korean}. This proves that our implementation in magma is as optimal as theirs. Even though our implementations in Magma are almost optimized, our findings are different than that of Cheon et.~al.~\cite{korean}. It would be nice, if we could have provided the exact reasons for this difference. However, since Cheon et.~al.~\cite{korean} were not forthcoming with their implementation details in their paper, we were unable to pinpoint the exact set of reasons.

As we expected the implementation of the original $r$-adding was much faster in C++. For $10^8$ iterations, it took about $820$ and $350$ seconds for arbitrary and sparse irreducible polynomial respectively as compared to $1210$ seconds for Magma. For the sparse polynomial, we were able to beat the time in ~\cite{korean}. However with an arbitrary polynomial we were not even close to their time of about $400$ seconds. We tried with two different programming languages -- NTL and Magma and many different implementations trying to speed up our implementations. All failed to reach their time.

At this point, we must say that our results in Magma and NTL are not in sync as is clear from Tables 2 and 3. We tried all we could to make respective implementation as effective as possible. A straightforward implementation was not fast. One central issue was, how the elements of a finite field is stored internally in these languages and how easy or difficult it is to make it a vector for tag computation. In the case of Magma, it was an unnecessary but unavoidable ``coercion'' that got in the way. In the case of NTL, it was an unnecessary but unavoidable ``type casting'' that got in the way. After going through these issues, we have serious doubts with these implementation times being considered as a \textbf{scientific evidence and any conclusion} that follows from that.

There are some advantages and disadvantages for using $r=4$ instead of $r=16$ in the modified walk. An advantage of $r=4$ is that the modified walk gives some speed-up and one can use large values of $l$ compared to that of $r=16$. The size of the table $\mathcal{M}_l$ is given by $\binom{l+r}{r}$. One disadvantage of $r=4$ is that the $4$-adding walk has large expected rho length and variance. This means that finding the collision $4$-adding method is quite uncertain. An advantage of $r=16$ is that the $16$-adding walk has small expected rho length and variance. This means that finding the collision using $16$-adding method is more certain compared to the $4$-adding walk. One disadvantage of $r=16$ is that the modified method gives almost no speed-up and one can not use large values of $l$. 

%\nocite{*}
\bibliography{paper}
\bibliographystyle{amsplain}
\end{document}